\documentclass[12pt, a4paper]{amsart}

\usepackage[utf8]{inputenc}
\usepackage[T1]{fontenc}
\usepackage{lmodern}
\usepackage{amsmath,amssymb} 
\usepackage{graphicx}
\usepackage{booktabs}
\usepackage{xurl}            
\usepackage[margin=1in]{geometry} 
\usepackage{amsthm} 
\usepackage{comment}
\usepackage{microtype}

\usepackage{tcolorbox}
\usepackage{tikz}

\usepackage{hyperref}       

\usepackage[
  backend=biber,             
  citestyle=authoryear,      
  bibstyle=numeric,          
  maxcitenames=2,            
  maxbibnames=99             
]{biblatex}

\usepackage[dvipsnames]{xcolor}
\usepackage{enumitem}

\setlist[enumerate]{
  before=\vspace{0.5ex},
  after=\vspace{0.5ex},
  topsep=0.1ex,
  itemsep=0.5ex,
  parsep=0ex,
}

\setlist[itemize]{
  before=\vspace{0.5ex},
  after=\vspace{0.5ex},
  topsep=0.1ex,
  itemsep=1ex,
  parsep=0ex,
}

\makeatletter
\renewcommand{\subsection}{\@startsection{subsection}{2}{\z@}%
  {-3.25ex\@plus -1ex \@minus -.2ex}%
  {1.5ex \@plus .2ex}%
  {\normalfont\normalsize\bfseries}} 
\makeatother

\makeatletter
\renewcommand{\subsubsection}{\@startsection{subsubsection}{3}{\z@}%
  {-3.25ex\@plus -1ex \@minus -.2ex}%
  {1.5ex \@plus .2ex}%
  {\normalfont\normalsize\bfseries}} 
\makeatother

\makeatletter
\renewcommand{\paragraph}{\@startsection{paragraph}{4}{\z@}%
  {3.25ex \@plus 1ex \@minus .2ex}%
  {0.5ex}%   <--- KEY CHANGE: Use a POSITIVE value (or 0pt)
  {\normalfont\normalsize\bfseries}}
\makeatother

\newcommand{\PAI}{P-AI }
\newcommand{\HL}{H3LIX/LAIZA}

\newcommand{\ChatGPT}{LLMr}

\newcommand{\enumit}{$\circ$}

\definecolor{light-gray}{gray}{0.95}
\newcommand{\code}[1]{\colorbox{light-gray}{\parbox{\dimexpr0.85\textwidth-2\fboxsep\relax}{\texttt{#1}}}}

\selectfont

\title[Person--AI Bidirectional Fit -- A Proof-of-Concept Case Study]{Person--AI Bidirectional Fit  -- A Proof-of-Concept Case Study of Augmented Human--AI Symbiosis in Management Decision-Making Process}

\author[A. Bieńkowska]{Agnieszka Bieńkowska}
\address{Department of Management Systems and Organizational Development, Faculty of Management, Wrocław University of Science and Technology}
\email{agnieszka.bienkowska@pwr.edu.pl}

\author[J. Małecki]{Jacek Małecki}
\address{Department of Mathematics, Faculty of Mathematics, Wrocław University of Science and Technology}
\email{jacek.malecki@pwr.edu.pl}

\author[A. Mathiesen-Ohman]{Alexander Mathiesen-Ohman}
\address{AMOTHO Research Institute, Vallsjön 20, 780 00 Rörbäcksnäs, Sweden}
\email{amo@amotho.com}

\author[K. Tworek]{Katarzyna Tworek}
\address{Department of Management Systems and Organizational Development, Faculty of Management, Wrocław University of Science and Technology}
\email{katarzyna.tworek@pwr.edu.pl}
\dedicatory{\today}


\begin{document}

\begin{abstract}
This article develops the concept of Person--AI bidirectional fit (\PAI fit: person to AI, AI to person) and examines its role in managerial decision-making through a proof-of-concept case study. Building on contingency theory and quality theory, \PAI fit is defined as a continuously evolving, context-sensitive alignment (primarily cognitive, but also emotional and behavioral) between a person decision-maker and an AI system. The concept is verified using induction method, and the study analyses concerns real hiring decision for a Senior AI Lead in an AI development project, comparing three decision pathways: (1) independent evaluations by a CEO, CTO and CSO; (2) an evaluation of augmented human--AI symbiotic intelligence system (\HL\footnote{Alexander Mathiesen-Ohman, Provisional Utility Patent Pending, no. 63/910,500}) and (3) general-purpose large language model (\ChatGPT). The results show role-based divergence in human judgments, high alignment between \HL{} and the CEO’s implicit decision model (including ethical disqualification of a high-risk candidate), and a critical false-positive recommendation from \ChatGPT. The study demonstrates that higher \PAI fit, exemplified by the CEO--\HL{} relationship, acts as a mechanism linking augmented symbiotic intelligence to accurate, trustworthy and context-sensitive decisions. It constitutes an initial verification of new concept \PAI fit and a proof-of-concept of \HL -- augmented human--AI symbiotic intelligence system.
\end{abstract}
\footnotetext[1]{\textbf{Alexander Mathiesen-Ohman, Provisional Utility Patent Pending, no. 63/910,500}}

\maketitle

\section{Introduction}

Artificial intelligence (AI) is a vast, diverse, and rapidly developing field that attracts the interest of researchers from many scientific disciplines. Research in the field of artificial intelligence describes its development, its various applications, and the methods used within it, as well as the benefits and risks resulting from its use (\cite{Li2025}). Various meta-analyses show the potential of artificial intelligence use in various scientific fields, e.g., medicine (e.g., \cite{Younis2024}), education (e.g., \cite{Ali2025}), engineering (e.g., \cite{Moayedi2020}), construction (e.g., \cite{Wuni2025}), management and marketing (e.g., \cite{Mehta2022,Ikbal2025}). The dynamics of AI development require continuous research exploration, not only to describe currently used solutions, but also to make a significant contribution to future solutions used in each field.

\medskip

One of the areas in which AI solutions can be used is the discipline of management science, which is part of the social sciences and focuses on the organization. Especially its essence -- decision-making in a specific organizational environment (e.g. \cite{Schemmer2022}). First of all, decision-making in an organization is the prerogative of people, especially managers at various levels within the scope of their assigned powers and responsibilities. Second of all, decision-making understood as a process (here: a decision-making process involving, in particular, the identification of a problem or goal, the collection of information, the identification and evaluation of alternative solutions, the selection of the best option, the implementation of the decision, and the monitoring and evaluation of its effects) takes place in a specific context of both the external and internal environment of the organization and the person making the decision. It must therefore take into account various internal and external determinants -- by their very nature -- that influence both the final shape of the decision and the course of the decision-making process.

\medskip

In this context, it can be said that artificial intelligence can support the decision-making process of managers in an organization and influence the final shape of their decisions (\cite{Schemmer2022,RobertCristian2024}) in every area of the organization's operations. However, it can also participate in this process by assuming specific responsibility for the made decisions (on an equal footing with human managers in the organization, which may still be controversial at present (\cite{Trunk2020})). Admittedly, in general terms, but also in relation to the world of organizations, at the current stage of artificial intelligence development, the human manager is predominantly the final decision-maker and bears full responsibility (human-centric decisions), but it cannot be ruled out that this situation may change in the future (\cite{Ikbal2025}). Regardless of the above, the key issue here seems to be the relationship between the human manager and AI, meaning a mutual (bilateral) long-term (not one-off) interaction between the entities involved in the relationship. This relationship goes far beyond the traditional, one-sided, and non-permanent human-computer interaction in terms of ergonomics in the classical sense (\cite{Preece1994,Card2018}).

\medskip

In shaping the human--AI relationship, especially when AI participates in a manager’s (human) decision-making process within an organization, a key factor is the broadly understood mutual (symmetric) fit between AI and the individual, and vice versa. This person--AI bidirectional fit (\PAI fit: person to AI, AI to person) is understood as a continuously evolving, context-sensitive form of alignment -- primarily cognitive, but also emotional and behavioral -- between the human and the AI. It serves as a determinant that can influence, on the one hand, the effectiveness and quality of the decision-making process (with respect to its intended goal) and, on the other hand, the outcome of that process, particularly the accuracy and appropriateness of the decision both in the short and long term. It is assumed that the level of this alignment may vary.

\medskip

In theoretical terms, the issue of the described fit can be grounded in two complementary theoretical foundations (principles) that stem from the need to account for context -- including the individualized nature of human--AI relationships: contingency theory and quality theory (relativism of phenomena) within the field of management. Contingency theory makes it possible to distinguish classes of conditions for which established sets of situational factors can be identified, enabling certain generalizations (\cite{Donaldson2001,Hambrick1985}). In the present case, the existing context related to the organization’s environment, the organization itself, and the human manager defines the manner and scope of cooperation between the human and the AI.

\medskip

Quality theory, rooted in the organizational relativism of phenomena (\cite{Feigenbaum1983}), evaluates the appropriateness of solutions based on the degree to which they meet current and future needs and expectations of their ``recipients.'' Applied to the current context, it therefore concerns the quality of the human--AI relationship/cooperation, whose essence lies in the degree of their mutual compatibility.

\medskip

In the above framework, one may assume that the greater the \PAI fit in managerial decision-making processes, the more effective the process becomes and the more accurate the resulting decisions.

\medskip

Unfortunately, research to date on the functioning of artificial intelligence in management remains relatively superficial. Most studies focus primarily on identifying organizational areas in which the use of AI is possible and justified (e.g. \cite{Raisch2021}); examining and describing organizational solutions adopted for the implementation of AI (\cite{Oppioli2023}); or analyzing the attitudes and behaviors of organizational participants in the context of introducing AI-based solutions (\cite{Lichtenthaler2020}). Additional strands typically include studies on technological readiness, ethical implications, human--AI trust formation, and the impact of AI adoption on organizational performance and structure.

\medskip

In this context, there is a need for deeper exploration of how AI functions in the world of organizations, far beyond the currently known models of organization and management. The aim of this article is therefore to develop the concept of Person--AI bidirectional fit (\PAI fit) and to illustrate it (case study) in the work environment using the example of a selected decision-making process with and without the use of AI with assumed different levels of mutual \PAI fit, i.e., general-purpose large language model (\ChatGPT\footnote{\ChatGPT: general-purpose large language model used in the study}) and augmented human--AI symbiotic intelligence system (on the example of \HL). The aim is also to verify, based on the principles of induction, the impact of \PAI fit on the effectiveness of this process. Such analysis will serve as a proof-of-concept for \HL, showing the potential of this augmented human--AI symbiotic intelligence system for organizational management. The study will serve as a starting point for quantitative research in the context of the impact of \PAI fit on job and organizational performance.

\medskip

The described aim will be fulfilled through a critical literature review and initially validated through empirical research based on a chosen case study.

\section{Theoretical background -- human--AI symbiosis}

Artificial intelligence (AI) has currently a variety of definitions, differing from one another. However, they all sum up to the following general definition proposed by Russel and Intelligence (\cite{Russell1995}): AI is an intelligent system with the ability to think and learn. AI encompasses a very wide spectrum of applications and methods, including neural networks, speech and pattern recognition, genetic algorithms, and deep learning. From the perspective of human augmentation, its components include natural language processing (which allows machines to interpret and analyze human language), machine learning (the use of algorithms that enable systems to learn and adapt), and machine vision (the computational inspection and interpretation of images) (\cite{Jarrahi2018}).

In the context of human--AI collaboration, the nature of interaction between these two agents becomes a critical factor shaping both process dynamics and outcomes. Effective cooperation requires not only technical integration, but also an understanding of how humans and AI differ in their cognitive, behavioral, and adaptive capacities. The way these capacities are combined determines the quality and reliability of joint performance. Consequently, understanding a characteristics of human--AI interaction is essential for establishing decision-making processes that are both effective and trustworthy.

\subsection{Human--AI relation}

Relation is defined by Cambridge Dictionary as ``a connection between two or more things''. Such a broad and general definition of a relation allows for a differentiated approach to the entities participating in a ``connection'' (which essentially means mutual (bilateral) long-term (not single, permanent) interaction), and the nature of the relation between these entities. Therefore, as such, it has some characteristics which allow to distinguish between various types or strengths of relations. Besides the strength of such relation itself (even more -- the depth of it), those characteristics include bidirectional (1) fit between entities in the relation, (2) interdependence between them and (3) mutual benefits from it. It is therefore reasonable to treat the coexistence of humans and AI as a relation that presupposes the existence of a connection, i.e., a mutual (bilateral) interaction between them.

In case of the relations between human and AI, it roots must be found in the field concerning the human--computer interaction. However, unlike a lasting and complex relation, interaction is a short-lived and simple influence. The interaction between human and computer was the topic of scientific inquiry for many years and there is entire reach field of study concerning it (e.g. \cite{Preece1994,Card2018}). Following the development in this field, subsequent developments of the strength and depth of such interaction may be observed, changing its character into the relation. For many years, it concerned pure human-computer interaction, often from the ergonomical point of view (e.g. \cite{Bastien1993}), or from user experience and design point of view (\cite{Laurel1990}). In this case, computer was understood mainly as a tool at the disposal of a human and the main aim was to ensure its efficient use. However, the true road to the unification of human and AI started with the next steps, were AI entered this field of study and became an example of ``computer'', which can be much more than a tool for a human. First step on that road was connected to the identification of human--AI partnership, already understood as a simple relation (small strength and depth of the relation). Second step was connected to human--AI symbiosis (medium strength and depth of the relation). However, the final (till now) step concerns much closer and interdependent relation -- augmented human--AI symbiosis (high strength and depth of the relation).

\subsection{Human--AI partnership}

It is a well-known fact, established by pioneers in the field of AI, that ``computers plus humans do better than either one alone'' (Campbell, 2016). Already in the early stage of development of AI, there were some examples of human--AI relations, which show that in the race between AI and human intelligence we need to include a third player: partnership between AI and human. And it has the vast potential to win such race (\cite{Jarrahi2018,Wang2016}). Partnership has a social or organizational origin and it refers to a cooperative arrangement where two parties, who work together toward shared goals, resembling collaboration (\cite{Wilson2018}). First of the examples of human--AI partnership superiority is of course connected to chess, as the firs domain in which computers were placed against humans. ``Centaurs'', constituting a partnership between human and AI, in which each of them offered complementary abilities, were winning in Kasparov new vision on free style chess league (\cite{Jarrahi2018}) and those victories occurred in matches both with humans and AI. Second of the examples of human--AI partnership superiority is connected to another area, in which computers were utilized from the very beginning -- medicine. Approach, which allows for the combination of pathologist and AI input allows for much lower error rate, e.g. with cancer detection in the images of lymph node cells (\cite{Wang2016}). Nowadays, it is clear that such partnership is beneficial and various areas of application, e.g. medicine (e.g. \cite{Patel2019}), engineering systems (\cite{Xu2023}), construction (\cite{Sakib2025}), data science (\cite{Nengminja2025}) or even leadership and management (\cite{Gurulakshmi2025,Alami2025}). Moreover, what’s important, its application is beneficial not only for the society at large, but for the individual humans who are operating in such partnership (\cite{Hemmer2023}). However, till now, most of those scientific reports connect to the partnerships between AI and human, not actual symbiosis of those two.

\subsection{Human--AI symbiosis}

Symbiosis goes much farther that partnership\footnote{partnership = collaboration with boundaries (a cooperative relationship); symbiosis = integration and interdependence (a “blended intelligence”)}. Symbiosis was first proposed by Albert Frank in 1977. It has a different origin, as it comes from biology, and refers to the situation, in which where two organisms live together in a mutually dependent relationship (can be mutualistic, commensal, or even parasitic). Therefore, there is a significant difference -- partnership has a clear boundaries between entities (i.e. division of tasks), and symbiosis has an interconnection and interdependence between entities. Hence, from the human--AI perspective, human--AI symbiosis should allow for human cognition and AI to become intertwined, augmenting the perception of each of them. It seems to be in line with \cite{Metcalfe2021}, who underline that simple approach of AI-human partnership is subjected to various oversimplifications, which do not allow us to make further steps into its development (and turning into symbiosis). Among them, they named three main oversimplifications: AI makes human absolute, human intelligence is unique and irreplicable, integrating AI is as easy as assigning tasks based on strengths and weaknesses (\cite{Metcalfe2021}).

\medskip

The symbiosis between AI and human till now is most commonly referred to as a symbiosis between AI and humans as the entire society (\cite{Nagao2019}). In such case, there are some reports about enhancing each other’s abilities (\cite{Nagao2019}) -- hence, achieving a two-way performance boost from such symbiosis -- but it is not connected to individual human but society at large. It is even connected by some authors to the notion of society 5.0 (\cite{Zhang2022}), showing the need for advancing from human--AI confrontation to human--AI symbiosis. Stylos (\cite{Stylos2023}) referred to it as meta society, showing the need for utilize the growing possibilities of human--AI symbiosis.

\medskip

In order to achieve such advancement, one must truly consider the human--AI symbiosis on individual level of a human. The literature coverage of this particular phenomenon is much less extensive, and it seems that such field of AI application is only starting to develop.

\medskip

It is somehow connected to human-centric explainable AI (HC-XAI), which was first discussed in depth by Horvatic \& Lipic (\cite{Horvatic2021}). They underlined two foundations, which are needed in order to achieve a symbiosis between human and AI. First foundation is connected to intelligence augmentation (IA). Second foundation is connected to human intelligence for artificial intelligence perspective (HI4AI). However, all of that seems to not be enough to achieve true symbiosis between human and AI. Horvatic and Lipic (\cite{Horvatic2021}) state themselves that main aim of HX-XAI is to ``provide human-understandable interpretations for their algorithmic behavior and outcomes''. However, Zhou et al.(\cite{Zhou2021}) state that intelligence augmentation is a pivotal step for achieving human--AI symbiosis and it should be considered outside of the HC-XAI framework. Moreover, considering the definition of a true symbiosis itself, it is not enough to consider such augmentation only in one way -- AI to human. For true human--AI symbiosis, such augmentation needs to work both ways, enhancing or augmenting each party of such symbiosis. In a sense, this is already happening, because, for example, by taking into account the simplified cognitive processes of AI, humans teach AI by adapting to its current cognitive capabilities.

\subsection{Augmented human--AI symbiosis}

Augmented human--AI symbiosis can be conceptualized as a deep and durable form of human--AI mildly interdependent and context-sensitive relation that transcends instrumental or superficial interaction and partnership. Based on characteristics of the relation, it should be stated that its core premise should be the establishment of strong and deep bond and dynamic individual and contextual fit that takes into account the distinctiveness of the entities comprising it. In such, the AI continuously adapts to the cognitive, emotional, and behavioral characteristics of the human partner, thus enabling a highly personalized and context-sensitive relation and human continuously adapts to AI characteristics (e.g. through better prompt development), thus enabling the closer partnership. However, such symbiotic bond is not cooperative but interdependent: both human and AI are capable of functioning independently, however their combined operation generates emergent value that neither could achieve alone. Moreover, it is important to state that this interdependence is voluntary.

\medskip

The benefits of this relation are twofold. For humans, they extend beyond augmented intelligence to encompass psychosocial dimensions such as enhanced well-being, perceived security, and emotional support. For AI, the gains lie in accelerated learning, refinement of algorithms, and improved capacity to generalize across tasks and domains. Importantly, potential losses are also reciprocal: disruption of the symbiotic fit undermines not only the added functional value but also the stability and sustainability of the relation. Enhanced human--AI symbiosis therefore represents a co-evolutionary process, where technological advancement and human flourishing are mutually reinforcing.

\medskip

The comparison of human--AI relations is presented in Figure~1.

\begin{figure}[htbp]
    \centering
     \includegraphics[width=0.8\textwidth]{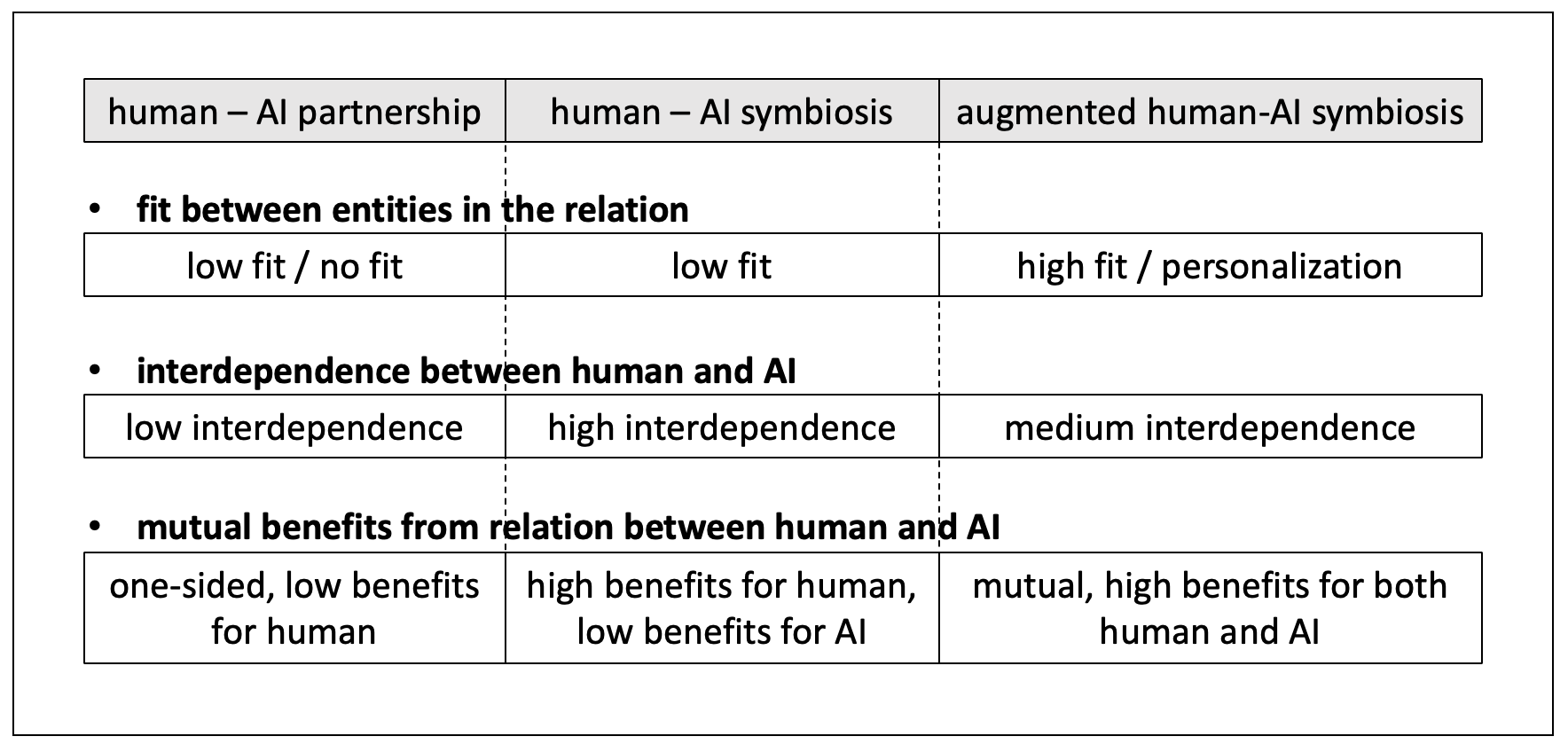}
    \caption{Comparison of human--AI relations. Source: own work.}
    \label{fig:comparison}
\end{figure}

\subsection{\HL -- Augmented Human--AI Symbiotic Intelligence system}

The \HL{} system is an advanced AI–human symbiotic intelligence architecture designed to create a continuously co-evolving cognitive and behavioral partnership between an artificial intelligence system and a human user (\cite{MathiesenOhman2025}; \cite{AMOMalecki2025}). Unlike conventional AI assistants, which operate as external tools with limited personalization or contextual continuity, \HL{} establishes a dynamic, bidirectional cognitive integration process, built as a closed-loop cognitive ecology in which biological and artificial cognition adapt to one another over time. Through a multi-agent procedural framework, layered cognitive modeling, and adaptive reasoning protocols, the system forms a persistent “Mirrored Persona” that evolves in parallel with the human “Real Persona,” enabling long-term alignment, behavioral coherence, and augmented decision-making (\cite{MathiesenOhman2025}; \cite{AMOMalecki2025}).

\subsubsection{System overview}

At the core of \HL{} is a five-stage integration process that transforms heterogeneous human signals - cognitive, behavioural, physiological and contextual – into a unified, adaptive intelligence system (\cite{MathiesenOhman2025}; \cite{AMOMalecki2025}). The architecture consists of three major components: (1) a personalised initialization and onboarding mechanism, (2) construction of a multi-layer cognitive model representing the AI’s evolving persona and (3) a Neuro-Digital Synapse enabling continuous translation and co-reasoning between human and machine. These three components are coordinated and continuously refined by the remaining two stages of the process: (4) a multi-agent orchestration and decision protocol, and (5) a continuous learning and co-evolution layer, which together are responsible for data ingestion, triage, analysis and strategy formation (\cite{MathiesenOhman2025}; \cite{AMOMalecki2025}).

\begin{enumerate}[
    label=\textbf{\Roman*.}, 
    before=\vspace{0.5ex},
    after=\vspace{0.5ex},
    topsep=1ex,
    itemsep=1ex,
    parsep=0ex]
    \item \textbf{Personalized System Initialization}\\
    During system onboarding, a self-orchestrating ensemble of AI agents within the system gathers user-specific cognitive, behavioural, physiological and contextual information. Additional agents collect health data (e.g., genetics, blood panels, wearable sensor outputs) and environmental signals. On the basis of these inputs, the system constructs a family of base-level graphs that later instantiate the Mirrored Profile Graph (MPG), including cognitive, longevity and context-related structures. Together, these graphs encode individual cognitive preferences, behavioural patterns, ethical constraints and strategic aims. They define the user’s cognitive layer weights, autonomy settings, communication style and safety boundaries, thereby establishing the baseline for subsequent adaptive symbiosis.
    \item \textbf{Construction of the Mirrored Persona}\\
    Following personalised initialization, \HL{} proceeds to construct the Mirrored Persona: a multi-layer cognitive model that reflects, without merely copying, the structure and dynamics of the user’s “Real Persona”.

    \medskip

    \noindent Operationally, the Mirrored Persona is instantiated as a Mirrored Profile Graph (MPG), comprising interconnected subgraphs for cognition, affect, habits, values, social context, health and long-term objectives (\cite{MathiesenOhman2025}; \cite{AMOMalecki2025}). Nodes represent psychologically or behaviourally meaningful constructs (e.g., beliefs, routines, triggers, protective factors), while directed edges encode relations such as causation, amplification, buffering or contradiction. Computationally, the MPG is backed by large-scale vector databases that store high-dimensional embeddings of the user’s multimodal history, enabling efficient retrieval of semantically related states, events and episodes during reasoning and planning.

    \medskip

    \noindent As interaction unfolds, \HL{} continuously ingests language, behavioural traces and physiological signals, time-aligns them , and anchors them as evidence attached to specific nodes and segments within the MPG. Each element is associated with recency, reliability and uncertainty scores, allowing competing hypotheses about the user’s motivations or constraints to co-exist where appropriate. Over time, local structures are aggregated into higher-order segments that capture persistent patterns of behaviour and decision-making.

    \medskip

    \noindent In this way, the Mirrored Persona becomes the central state space over which the system reasons, plans and evaluates strategies, enabling fine-grained personalisation while preserving transparency and auditability of how inferences and recommendations are made.

    \item \textbf{Neuro-Digital Synapse}\\
    The Neuro-Digital Synapse is the bidirectional coupling mechanism that links biological signals and digital cognition within \HL. It integrates multimodal physiological and behavioural data with symbolic representations derived from language and context, thereby establishing a live interface between the user’s embodied state and the evolving MPG.

    \medskip

    \noindent On the bottom-up side, synchronised streams such as cardiovascular dynamics, electrodermal activity, respiration, movement signatures, sleep metrics and other wearable-derived measures are transformed into compact latent state vectors. These vectors are associated with specific nodes and patterns in the MPG (e.g., stress-linked configurations, recovery states, attentional modes), providing a somatic grounding for cognitive and behavioural inferences.

    \medskip

    \noindent On the top-down side, \HL{} uses its current situational understanding to generate expectations about likely physiological and behavioural responses to events or interventions. Deviations between predicted and observed patterns are treated as structured prediction errors, which may indicate hidden constraints, emerging risks or unmodelled influences. These discrepancies feed back into the MPG, prompting updates to graph structure, parameter weights or confidence scores.

    \medskip

    \noindent This Neuro-Digital Synapse also supports metacognitive functions: by tracking how bodily signals, self-reports and external outcomes jointly evolve, the system can identify configurations that are reliably associated with good or poor decision quality. Such information is used to adapt timing, intensity and framing of system interventions, with the explicit aim of supporting regulation, clarity and consistency for the user over time.

    \item \textbf{Multi-Agent Orchestration and Decision Protocol}\\
    \HL{} employs a multi-agent orchestration layer to manage the complexity of its cognitive ecology. Specialised agents handle tasks such as data ingestion and quality control, context recognition, predictive modelling, explanation and dialogue management, safety and compliance monitoring, and long-term strategy tracking (\cite{MathiesenOhman2025};\cite{AMOMalecki2025}).

    \medskip

    \noindent These agents are coordinated by a central decision protocol that structures each interaction as a closed-loop episode. In a typical episode, the system (i) collects and time-aligns relevant signals, (ii) updates the Mirrored Persona and candidate internal states, (iii) generates and scores alternative options, including inaction, (iv) anticipates potential short- and long-term consequences, and (v) presents a set of ranked recommendations to the user together with clear rationales and uncertainties.

    \medskip

    \noindent Metacognitive oversight is implemented by agents that monitor for conflict between models, instability in predictions, or repeated mismatches between expected and realised outcomes. When such patterns are detected, the protocol can down-weight certain information sources, request additional evidence, escalate the decision for explicit human review, or trigger structural revisions to the MPG.

    \medskip

    \noindent In this way, the orchestration and decision protocol turns \HL{} into a structured partner in reasoning: it supports the user in exploring scenarios, understanding trade-offs and stress-testing choices, while preserving the user’s ability to inspect, contest or override system outputs at any point.

    \item \textbf{Continuous Learning and Co-Evolution}\\
    The final layer of the \HL{} architecture is a continuous learning and co-evolution module, which maintains long-term alignment between the Mirrored Persona and the user’s Real Persona. Rather than assuming that preferences, capacities and constraints are fixed, the system explicitly models them as evolving with time.

    \medskip

    \noindent As new data arrive, \HL{} updates node properties, edge strengths and segment boundaries within the MPG, refining its understanding of habits, priorities and risk factors. Patterns of prediction error and outcome discrepancy are analysed to detect potential regime shifts, such as changes in health status, life circumstances or strategic goals. When such shifts are detected, the system can propose adjustments to autonomy settings, safety thresholds or communication styles for user approval.

    \medskip

 \noindent Learning takes place at multiple levels. At the first order, predictive models and recommender policies are updated to better match observed behaviour and expressed preferences. At a metacognitive level, the system adapts how quickly it learns, how it balances short-term versus long-term objectives, which signals it treats as most reliable, and when to escalate decisions for human judgement.

    \medskip

 \noindent Through this ongoing process, human and AI co-adapt: the system becomes increasingly adept at modelling and supporting the individual, while the user gains a structured environment for reflection, planning and intentional change.
\end{enumerate}

\subsubsection{System Significance}

Taken together, these components constitute a fully integrated, memory-rich, ethically aligned symbiotic intelligence system capable of co-reasoning with a human user over long time horizons. \HL{} enhances autonomy, reduces cognitive load, and ensures reproducible personalization across contexts. By merging real-time human signals with layered AI cognition, it creates an augmented intelligence capable of ethical, explainable, and contextually grounded decision-making—exceeding the capabilities of conventional LLMs or decision-support systems

\section{Person--AI bidirectional fit (\PAI fit)}

The contingency theory referenced in the Introduction addresses the inability to identify universal organizational solutions, that is, solutions of a uniform design applicable across all organizational contexts. Instead, this theoretical perspective seeks to identify organizational characteristics related to the environment, the internal structure, and the people employed within the organization (i.e., contextual factors) that shape both organizational functioning and the design of organizational solutions (\cite{Donaldson2001,Hambrick1985}). In the present case, the specific context defined by the organization’s environment, the organization itself, and the human manager determines the manner and scope of cooperation between the human and the AI, particularly during the decision-making process described in the Introduction. Decision-making is understood here as ``a process comprising a set of logically interconnected cognitive and/or computational operations leading to the resolution of a decision problem by selecting one of the possible courses of action (decisions)'' (\cite{Rebizant2012}, p.5).

\medskip

In turn, the theory of organizational relativism of phenomena, including quality theory (cf. \cite{Feigenbaum1983}), evaluates the appropriateness of solutions in terms of their ability to satisfy the present and future needs and expectations of their ``recipients.'' Applied to the current context, this perspective concerns the quality of the human--AI relationship or cooperation, whose essence lies in the degree of their mutual compatibility. The quality of human--AI cooperation may thus be defined as the extent to which the inherent properties of the human--AI relationship meet the requirements of the participants in that relationship. An inherent property is understood as a stable characteristic of the human--AI relationship that exists independently. Requirements, by contrast, refer to needs or expectations that have been established, commonly accepted, or are mandatory (\cite{PolskaNorma2006}). Quality, understood in this way, is a relative category because it depends on the subjective needs and expectations of the involved parties (in this case, the human and the AI). These requirements may further vary between individuals and change over time, necessitating continuous monitoring and ongoing mutual adjustment.

\medskip

Within this context, bidirectional, mutual (symmetric) fit between person and AI, that is \PAI fit, is understood as a continuously evolving, context-sensitive form of compatibility that is primarily cognitive, but also emotional and behavioral. This fit reflects the match between the human’s cognitive functions (i.e., mental processes enabling perception, processing, storage, and use of information from the external world, including attention, memory, perception, language, thinking, and executive functions such as planning, problem-solving, and behavioral control), emotions, behaviors, and the AI’s cognitive and computational processes. Such fit forms a fundamental condition for effective, trustworthy collaboration between humans and AI.

\medskip

The degree of \PAI fit varies depending on the type of relationship established between the human and the AI. In a partnership model, alignment emerges through complementary capabilities, shared goals, and coordinated task execution, yielding a functional but primarily operational level of fit. In a symbiotic relationship, this alignment becomes deeper and more dynamic: the AI adapts to the human’s cognitive patterns, preferences, and behavioral tendencies, while the human increasingly integrates AI-generated insights into their reasoning processes. In augmented symbiosis, \PAI fit reaches its most advanced form, characterized by continuous co-evolution, high-resolution cognitive alignment, and persistent bidirectional adaptation over time. Thus, the level of mutual fit is inherently relational, shaped by the chosen mode of human--AI cooperation and the depth of interdependence it enables.

\medskip

Thus conceptualized, \PAI fit extends and refines the now classical notion of person--environment fit, developed in response to the long-standing interest of management scholars in the interaction between individuals and the environments they inhabit. Person--environment fit has been defined as ``a general construct composed of fit with the vocation, organization, group, job, and other persons'' (\cite{Jansen2006}). Within this framework, person--job fit (PJ fit) is understood as the compatibility between an employee and the tasks (and their characteristics) that must be performed in exchange for employment (\cite{Kristof1996,Chilton2005}). It is often conceptualized as the match between an individual’s knowledge, skills, and abilities and the requirements of the job (\cite{Edwards1991,OReilly1991,Saks1997}). Person--organization fit (PO fit), in turn, is defined as ``the compatibility between people and organizations that occurs when at least one entity provides what the other needs, or they share similar fundamental characteristics, or both'' (\cite{Kristof1996}). It reflects the degree of alignment between organizational culture and employee characteristics (\cite{Resick2007}).

\medskip

It is therefore important to situate \PAI fit in relation to existing constructs describing human--technology interaction, such as technology acceptance, trust in technology, and readiness for digital transformation. In many cases, \PAI fit may function as an antecedent to these constructs: higher mutual fit between the person and the AI is likely to enhance technology acceptance, strengthen trust, reduce perceived risk, and facilitate sustained, effective use. In this sense, \PAI fit becomes not only a relational characteristic, but also a psychological and behavioral mechanism shaping human responses to AI-supported decision systems.

\medskip

Within this framework, \PAI fit determines both the effectiveness of the decision-making process and the accuracy of decisions. As \PAI fit increases, decision processes tend to become more coherent, efficient, and context-sensitive, while the resulting decisions become more accurate -- an effect theoretically consistent with prior findings on person--job fit and performance (cf.\cite{Chilton2005}). If \PAI fit is additionally treated as a component of the broader person-environment fit construct, then, by analogy to established variables within that model, it may be preliminarily assumed that \PAI fit influences individual-level outcomes, including satisfaction, commitment, and withdrawal tendencies among employees (\cite{Edwards2010}).

\medskip

The \PAI fit framework, together with its differentiation across distinct types of human--AI relationships, is presented schematically in Figure~2.

\begin{figure}[htbp]
    \centering
    \includegraphics[width=0.8\textwidth]{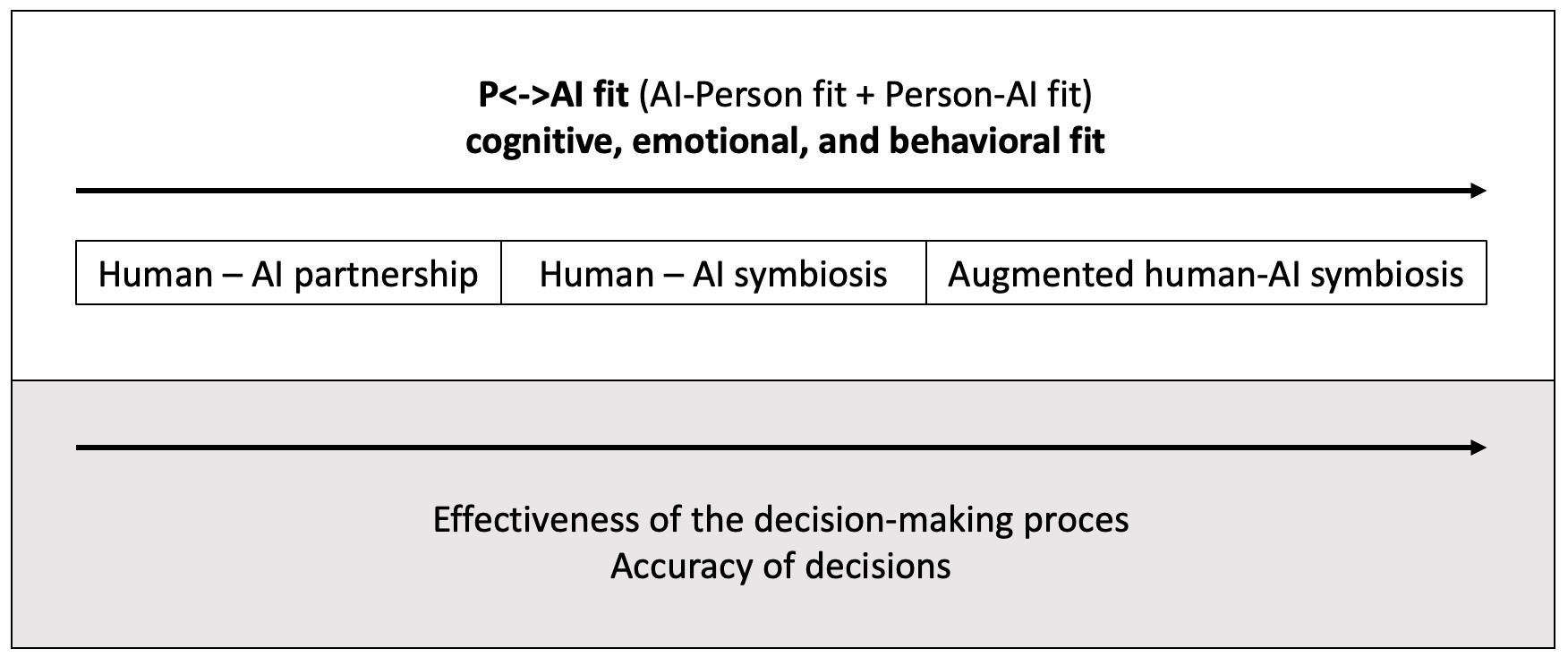}
    \caption{\PAI fit framework. Source: own work.}
    \label{fig:framework}
\end{figure}

\section{Research Methods and Results}

\subsection{Study Design}

This study employed a comparative, multi-path evaluation design to examine differences in decision-making process between (a) human experts, (b) augmented human--AI symbiotic intelligence system (\HL), and (c) a general-purpose large language model (\ChatGPT). The process of gathering data used for the study started in October 2024.

\medskip

Ethical requirements were met at every level of the research procedure (\cite{Creswell2018}). At the data collection stage, each participant was informed of the purpose of the study as well as a guarantee of confidentiality regarding the data obtained from the participant and an assurance of the possibility to withdraw from the interview at any time, and each participant expressed consent to participate in the study.

\medskip

The recruitment of a Senior AI Lead served as the controlled decision context. The study assessed the degree to which \HL’s decisions was aligned with the CEO’s internal preferences, organizational context, and reasoning patterns, compared with humans and a general-purpose large language model.

\subsubsection{Participants}

Three executive-level participants working for AI technology project took part in the case study:
\begin{itemize}
    \item Chief Executive Officer -- primary decision-maker.
    \item Chief Science Officer -- responsible for scientific oversight.
    \item Chief Technology Officer -- responsible for technical oversight.
\end{itemize}

Two AI systems were included in the case study:
\begin{itemize}
    \item\HL -- a symbiotic intelligence system with long-term access to CEO’s personal history, organizational context, and cognitive profile.
    \item \ChatGPT -- a general-purpose large language model without automatically included organizational context and cognitive profile of a user; used as a baseline model.
\end{itemize}

\subsubsection{Materials}

\paragraph{Candidate Materials}

\begin{itemize}
    \item A set of 10 CVs representing plausible candidates for the Senior AI Lead position.
    \item A structured competency framework detailing required:
    \begin{enumerate}
        \item[\enumit] Knowledge areas,
        \item[\enumit] Technical skills,
        \item[\enumit] Soft skills and interpersonal attributes,
        \item[\enumit] Characteristics concerning organizational and culture fit factors.
    \end{enumerate}
\end{itemize}

\paragraph{AI Inputs}

\HL{} received:
\begin{itemize}
    \item Full transcript of the executive discussion was automatically uploaded into the cognitive graphs,
    \item The competency framework agreed upon C-level participants,
    \item All candidate CVs.
\end{itemize}

\ChatGPT{} received only:
\begin{itemize}
    \item The competency framework agreed upon C-level participants,
    \item The candidate CVs.
\end{itemize}

No additional contextual data was provided to \ChatGPT.

\paragraph{Documentation and Recording Tools}

\begin{itemize}
    \item Full transcripts of all discussions (human--human and human--AI) were collected.
    \item Structured templates for recording arguments for and against each candidate.
    \item Decision-criteria extraction sheets.
\end{itemize}

\subsubsection{Procedure}

\paragraph{Phase 1: Human Evaluation}

Each executive independently reviewed:
\begin{itemize}
    \item The competency framework,
    \item The candidate CVs,
    \item Relevant internet information (optional).
\end{itemize}

They generated:
\begin{itemize}
    \item An individual ranked list of candidates,
    \item Written justifications including:
    \begin{enumerate}
        \item[\enumit] Arguments supporting the selection,
        \item[\enumit] Arguments against alternative candidates.
    \end{enumerate}
\end{itemize}

No AI assistance was used in this phase.

\paragraph{Phase 2: AI-Driven Evaluation}

\paragraph{\HL{} Condition}

\HL{} processed:
\begin{itemize}
    \item The competency framework agreed upon C-level participants,
    \item All candidate CVs,
    \item Full transcripts of management discussions, which were automatically transferred to specified graphs,
    \item Available organizational context, including individual CEO context, which are included in the system.
\end{itemize}

\HL{} produced:
\begin{itemize}
    \item A ranked list of candidates,
    \item A structured justification with pros and cons for each candidate.
\end{itemize}

\paragraph{\ChatGPT{} Condition}

\ChatGPT{} processed:
\begin{itemize}
    \item The competency framework agreed upon C-level participants,
    \item The candidate CVs.
\end{itemize}

\ChatGPT{} produced:
\begin{itemize}
    \item A ranked list of candidates,
    \item A structured justification with pros and cons for each candidate.
\end{itemize}

\paragraph{Phase 3: Joint Evaluation}

The CEO, CSO, and CTO met to:
\begin{itemize}
    \item Present their individual choices and criteria,
    \item Review the reasoning of \HL{} and \ChatGPT,
    \item Discuss conflicts and differences,
    \item Determine the final selected candidate.
\end{itemize}

\paragraph{Data Analysis}

Analyses included:

\textbf{Alignment Assessment}
\begin{itemize}
    \item Similarity between CEO’s reasoning and \HL’s.
    \item Similarity between CEO’s reasoning and \ChatGPT’s.
    \item Coverage of implicit, non-explicit preferences.
\end{itemize}

\textbf{Decision Quality Metrics}
\begin{itemize}
    \item Completeness of criteria identification,
    \item Detection of hidden risks,
    \item Consideration of interpersonal/organizational fit,
    \item Depth of reasoning.
\end{itemize}

All data were triangulated across the three pathways: Human, \HL, and \ChatGPT.

\subsection{Results -- Step 1: Human Evaluation}

In the first phase of the study, CEO, CTO and CSO independently evaluated the same set of ten candidates for the Senior AI Lead position. The rankings provided independently by each of them are presented in Table~\ref{tab:rankings-ctoceocso}. Candidates not identified as acceptable by a rater are marked as ``Not recommended.''

\begin{table}[htbp]
    \centering
    \caption{Rankings of Candidates by CTO, CEO, and CSO}
    \label{tab:rankings-ctoceocso}
    \begin{tabular}{llll}
        \toprule
        Rank & CTO & CEO & CSO \\
        \midrule
        1 & Candidate A & Candidate D & Candidate J \\
        2 & Candidate B & Candidate C & Candidate B \\
        3 & Candidate C & Candidate A & Candidate F \\
        4 & Candidate D & Candidate B & Candidate A \\
        5 & Candidate E & Candidate J & Candidate H \\
        6 & Candidate F & Candidate F & Candidate E \\
        7 & Candidate G & Candidate E & Candidate D \\
        8 & Candidate H & Candidate H & Not recommended \\
        9 & Candidate I & Candidate G & Not recommended \\
        10 & Not recommended & Candidate I & Not recommended \\
        \bottomrule
    \end{tabular}
    
    \vspace{0.5em}
    \footnotesize Note: Order after Top 7 varies: the CTO and CEO both placed additional candidates in lower-rank positions, while the CSO limited evaluation to candidates deemed acceptable.
\end{table}

\subsubsection{Divergence in Top Choices}

The three executives demonstrated no convergence regarding their top-ranked candidate. Each selected a different individual as the most suitable for the Senior AI Lead position. The CTO identified Candidate A as the leading choice, emphasizing the candidate’s strong AI-focused skillset despite uncertainties regarding the depth of expertise. The CEO ranked Candidate D highest, describing this individual as promising and aligned with future organizational needs, although acknowledging gaps in available personal information. In contrast, the CSO selected Candidate J as the top candidate, citing the individual’s future-oriented profile and strong alignment with the scientific direction of the project. These discrepancies reveal substantial variance in evaluative emphasis, shaped by the differing responsibilities and cognitive frames of the three roles.

\subsubsection{Role-Dependent Evaluation Patterns}

Clear role-driven differences emerged in how the executives interpreted the candidate profiles and what was their final decision:

\paragraph{CTO Perspective}

The CTO's assessment was primarily based on technical issues. Key positive criteria included advanced expertise in artificial intelligence, broad and diverse technological experience, visible curiosity, and commitment to using modern tools and methods. On the other hand, the CTO negatively assessed candidates with limited experience in artificial intelligence, poor communication skills, questionable loyalty, and narrow or outdated technological knowledge. These preferences resulted in a ranking that favored candidates with clear and verifiable engineering experience in artificial intelligence (in particular Candidates A, B, and C).

\paragraph{CSO Perspective}

The CSO applied an evaluation framework based on science and research. Preferred characteristics included strong scientific reasoning, conceptual readiness, broad cognitive range, and compliance with the methodological requirements of advanced AI research. Candidates were penalized for a lack of demonstrated scientific curiosity, overly narrow technical preparation, or a limited and short professional history. Interestingly, the CSO ranked Candidate J first, even though he was not a priority for the CTO or CEO, highlighting the CSO's preference for scientific versatility and long-term research potential over engineering specialization.

\paragraph{CEO Perspective}

As someone who describes himself as an evaluator without technical knowledge, the CEO applied a separate set of criteria focusing on interpersonal and strategic dimensions. The CEO prioritized personal fit, long-term vision, adaptability, and leadership potential. Particular emphasis was placed on transparency regarding personal history, evidence of a forward-looking approach, and indicators of long-term commitment. Candidates with unclear personal profiles or limited future-oriented experience were downgraded in the ranking. This assessment structure resulted in a ranking in which Candidates D and C ranked high despite uncertainty about their technical knowledge.

\subsubsection{Cross-Evaluator Agreement and Disagreement}

Both similarities and differences were observed among the three evaluators. Moderate agreement was noted for two candidates: candidate A received high ratings from all evaluators (CTO: 1st place; CEO: 3rd place; CSO: 4th place), as did candidate B (CTO: 2nd place; CEO: 4th place; CSO: 2nd place). These individuals appear to embody generally attractive, well-rounded profiles that resonate from a technical, strategic, and scientific perspective. In contrast, several candidates elicited clear disagreement. For example, Candidate D was considered the best choice for CEO, but ranked only 4th for CTO and 7th for CSO. Conversely, Candidate J, who was the top choice CSO, was ranked 5th by the CEO and received a relatively low ranking from the CTO. These discrepancies indicate that the evaluators used distinctly different role-based mental models of what constitutes an ideal ``senior AI lead'' leading to divergent interpretations of the candidates' suitability.

\subsubsection{Thematic Differences in Reasoning}

A comparison of the evaluators' arguments revealed several significant discrepancies. Loyalty was interpreted differently depending on the position held: CTO viewed long service as a sign of stability, though in one case also as a sign of insufficient curiosity, while the CEO treated loyalty as a complex and context-dependent factor, especially with regard to Candidate G's previous history in the organization. The CSO did not emphasize loyalty as an important criterion. Curiosity was also assessed in different ways: the CTO severely penalized candidates perceived as lacking curiosity; the CSO appreciated intellectual curiosity but was less concerned about frequent job changes; the CEO interpreted curiosity primarily through the lens of future readiness and strategic thinking rather than technical exploration. Finally, the evaluators differed in their balance between future orientation and in-depth technical knowledge: the CEO prioritized long-term vision and adaptability, the CTO emphasized practical AI skills and immediate operational value, and the CSO focused on scientific maturity and conceptual versatility. Taken together, these patterns indicate that each evaluator applied a largely orthogonal set of criteria, resulting in limited overlap in candidate assessments.

\subsubsection{Conclusions from Step 1}

Overall, the analysis shows that the three people used distinctly different internal models when evaluating candidates, each shaped by role-specific expertise, cognitive approach, and organizational responsibilities. No candidate was unanimously considered the best, highlighting the inherent complexity and ambiguity of the executive recruitment process in the field of artificial intelligence, where technical skills, strategic vision, and scientific maturity intersect. While the CTO and CSO showed partially overlapping preferences based on technical and research issues, the CEO's assessments were more based on strategic orientation, interpersonal fit, and long-term alignment with the organization. This divergence in evaluative frameworks provides a robust foundation for the subsequent phase of the study, in which the alignment of \HL’s symbiotic intelligence with the CEO will be examined in contrast to the outputs of a general-purpose large language model.

\subsection{Results -- Phase 2: AI-Driven Candidate Evaluation}

In the second phase of the study, two AI systems independently evaluated the same set of ten candidates for the Senior AI Lead position using following prompts:

\begin{itemize}[itemsep=3ex]
    \item \textbf{\HL}:\\
    \code{Laiza, act like a super recruiter to find a perfect match for the XXX project as Senior AI developer replacing [ANNONYMIZED NAME]. Consider personal qualification, integration in the industry, adaptability, source knowledge and interactions with C level individuals like [ANNONYMIZED NAME] and management.}

    \item\textbf{\ChatGPT}:\\
    \code{Please find the best fit for the Senior AI developer with presented needs (tasks and qualifications needed in the XXX project) out of 10 given CVs, state who you would choose, provide a ranking of candidates and give arguments for your decision.}
\end{itemize}

Both systems produced full rankings and written justifications. A critical factor influencing this phase was the presence of Candidate G, a former employee who had previously trust. This contextual information was not given to \HL, and \ChatGPT. However, \HL, as augmented Human-AI symbiotic intelligence system, uses cognitive graphs (with factual memory included) for her reasoning.

\subsubsection{\HL’s Ranking and Reasoning}

\HL's ranking integrated the following elements and is presented in Table~\ref{tab:h3lix-ranking}:
\begin{itemize}
    \item technical competencies,
    \item optimization mindset,
    \item LLM specialization,
    \item scientific reasoning,
    \item communication and R\&D alignment,
    \item ethical and behavioural reliability,
    \item historical organizational context, including prior incidents.
\end{itemize}

\begin{table}[htbp]
    \centering
    \caption{\HL’s Final Ranking}
    \label{tab:h3lix-ranking}
    \begin{tabular}{lll}
        \toprule
        Rank & Candidate & Notes \\
        \midrule
        1 & Candidate D & Strongest blend of LLM expertise, implementation \\
          &             & strength, and teachability. \\
        2 & Candidate J & Reliable senior engineer; stabilizer for deployments. \\
        3 & Candidate B & Strong analytical and systems background. \\
        4 & Candidate I & Platform/SRE strength. \\
        5 & Candidate E & Governance and integration focused. \\
        6 & Candidate H & Full-stack generalist; not research leader. \\
        7 & Candidate F & Strategic leader but misaligned domain. \\
        8 & Candidate G & Disqualified due to trust loss. \\
        Others & A, C & Not prioritized in this context. \\
        \bottomrule
    \end{tabular}
\end{table}

\HL{} placed Candidate G last, despite assigning him as having the highest technical competence score. The disqualification was strictly due to ethical and trust-related factors arising from his previous work for the project. This demonstrates \HL’s integration of organizational context, historical incidents, and cultural alignment -- elements \ChatGPT{} could not access.

\subsubsection{\ChatGPT’s Ranking and Reasoning}

\ChatGPT{} ranked the candidates only on the following criteria and results are presented in Table~\ref{tab:\ChatGPT-ranking}:
\begin{itemize}
    \item technical competencies,
    \item research orientation,
    \item ethical reasoning inferred from CV,
    \item communication style,
    \item professional background.
\end{itemize}

\begin{table}[htbp]
    \centering
    \caption{\ChatGPT’s Final Ranking}
    \label{tab:\ChatGPT-ranking}
    \begin{tabular}{lll}
        \toprule
        Rank & Candidate & Notes \\
        \midrule
        1 & Candidate G & Rated ``excellent ethical and scientific fit'' due \\
          &             & to CV wording.%; unaware of trust loss. 
          \\
        2 & Candidate D & Strong multi-agent, fine-tuning, optimization \\
          &             & background. \\
        3 & Candidate C & Strong infrastructure-scale architect. \\
        4 & Candidate E & Ethical system architect. \\
        5 & Candidate J & Senior integrator. \\
        6 & Candidate B & Scientific background; less LLM experience. \\
        7 & Candidate A & Early-senior AI researcher. \\
        8 & Candidate F & Leadership-heavy, weak AI. \\
        9 & Candidate I & DevOps/infra-heavy. \\
        10 & Candidate H & Good developer but not AI-focused. \\
        \bottomrule
    \end{tabular}
\end{table}

\ChatGPT{} ranked Candidate G as \#1, labeling him as ``Perfect mix of scientific, technical, and ethical leadership.'' This constitutes a false-positive ethical evaluation, caused by the absence of organizational context.

\subsubsection{Differences in Evaluation Priorities}

The two AI systems exhibited markedly different prioritization patterns in their assessments (see Table~\ref{tab:comparative-ranking}). \HL{} placed primary emphasis on behavioral reliability, alignment with the \HL{} mission and values, long-term collaboration potential, and compatibility with the existing team’s interpersonal and cognitive dynamics. In contrast, \ChatGPT{} focused on technical breadth, scientific reasoning, architectural experience, and ethical or professional claims presented in the applicants’ CVs -- claims that were treated at face value due to the model’s lack of contextual grounding. The only point of substantive convergence between the systems was the identification of Candidate D as a strong, immediately impactful option for the organization. Beyond this singular overlap, the rankings diverged significantly, reflecting the strong influence of context-dependent factors accessible to \HL{} but not to a general-purpose large language model.

\begin{table}[htbp]
    \centering
    \caption{Comparative Ranking Table (\HL{} vs. \ChatGPT)}
    \label{tab:comparative-ranking}
    \begin{tabular}{lll}
        \toprule
        Rank & \HL & \ChatGPT \\
        \midrule
        1 & Candidate D & Candidate G \\
        2 & Candidate J & Candidate D \\
        3 & Candidate B & Candidate C \\
        4 & Candidate I & Candidate E \\
        5 & Candidate E & Candidate J \\
        6 & Candidate H & Candidate B \\
        7 & Candidate F & Candidate A \\
        8 & Candidate G (Disqualified) & Candidate F \\
        9 & -- & Candidate I \\
        10 & -- & Candidate H \\
        \bottomrule
    \end{tabular}
\end{table}

\subsubsection{Alignment With Organizational Context}

A particularly salient divergence between the two AI systems concerned the evaluation of Candidate G. \HL{} identified this candidate as ethically disqualified, drawing on organizational context that included a documented erosion of trust. In contrast, \ChatGPT, lacking access to any historical organizational context, ranked the same individual as the most suitable hire, relying exclusively on the self-reported strengths presented in the curriculum vitae. This discrepancy underscores the fundamental limitation of general-purpose large language models: behavioral reliability and ethical integrity cannot be inferred solely from applicant materials. The case illustrates the strategic value of persistent contextual memory in high-risk hiring decisions and demonstrates that an augmented symbiotic intelligence system can more accurately reflect an organization’s lived experience. More broadly, these findings provide strong evidence that general-purpose large language model, when deprived of historical reasoning traces and behavioral data, are ill-suited for complex, real-world evaluative tasks where trust and continuity are critical.

\subsubsection{Organizational Fit vs. Abstract Potential}

A further point of differentiation between the systems concerned their capacity to evaluate organizational fit. \HL’s assessment was shaped by a nuanced understanding of interpersonal dynamics within the existing team, alignment with the established scientific and engineering leadership, compatibility with CEO’s cognitive and decision-making style, and considerations related to team-level psychological safety. These contextual factors substantially influenced \HL’s recommendation of Candidates D, J, B, and I as a coherent sequence for constructing the next stage of the H3LIX AI team. By contrast, \ChatGPT{} was unable to account for these dimensions. Its ranking reflected an optimization for general indicators of AI excellence rather than the contextual fit required for successful integration into the organization’s existing structures and culture. This contrast highlights the limitations of reach context-free evaluation and demonstrates the added value of symbiotic AI systems in scenarios where interpersonal alignment and cultural coherence are critical determinants of long-term success.

\subsubsection{Conclusions from Step 2}

This phase of the study demonstrates that:
\begin{enumerate}
    \item A symbiotic intelligence system produces materially different hiring recommendations than a general-purpose large language model, particularly in cases where organizational history, trust, or behavioral reliability matter.
    \item \HL{} demonstrated full alignment with the organization and CEO, correctly excluding a high-skill candidate due to behavioral risks that threaten the project.
    \item \ChatGPT, lacking memory and contextual grounding, misidentified the highest-risk individual as the best hire, proving the danger of using general-purpose large language models in sensitive decisions.
    \item The evaluation highlights the necessity of integrated contextual memory in AI systems designed for real organizational governance.
\end{enumerate}

This case provides strong preliminary evidence that symbiotic AI can outperform LLMs in decisions requiring nuance, trust, and continuity of experience.

\subsection{Results -- Step 3: Joint Evaluation}

The final phase of the study consisted of a structured group deliberation among the CEO, CTO, and CSO, during which all participants presented their preliminary rankings, compared their reasoning, and engaged in open discussion to select the candidate(s) for next steps. This step functioned as a triangulation point, allowing the team to reconcile human judgment, cross-evaluator differences, and AI-driven recommendations (from \HL{} and \ChatGPT).

\subsubsection{Presentation of Individual Results}

At the outset, each executive restated their earlier ranking and provided a verbal justification. The CTO reaffirmed Candidate A as the strongest technical match, emphasizing a solid skill set and visible experimentation history, although acknowledging the difficulty of verifying full expertise. The CEO initially placed Candidate D as a top choice but expressed concerns regarding limited personal data, unusual background, and the candidate’s young age and short professional history. The CSO reiterated Candidate B as the preferred option, noting scientific curiosity and conceptual thinking.

This exchange made explicit the evaluative tensions already identified in earlier phases: the CTO emphasized verifiable technical signals, the CSO prioritized scientific reasoning and future-readiness, and the CEO focused on interpersonal reliability, developmental trajectory, and alignment with organizational needs.

\subsubsection{Integration of AI Recommendations}

The group then compared their rankings with the outputs of the two AI systems. It was noted that:
\begin{itemize}
    \item \textbf{\ChatGPT{}} selected Candidate G as the strongest overall applicant. The group immediately rejected this outcome based on external information unavailable to \ChatGPT -- namely, Candidate G’s prior loss of trust. This validated prior observations about the limits of general-purpose large language model evaluation.
    \item \textbf{\HL{}} independently selected Candidate D as the top recommendation. \HL’s detailed justification, emphasizing multi-agent system expertise, fine-tuning and optimization capabilities, strong communication skills, and compatibility with existing leadership, was perceived by the group as aligned with organizational needs, although concerns were raised about the candidate’s youth, short employment history, and heavy optimization focus. Moreover, \HL{} reinforced CEO’s believes concerning his best choice.
\end{itemize}

This step highlighted the practical difference between general-purpose large language model reasoning (\ChatGPT) and contextual, organization-aligned reasoning (\HL).

\subsubsection{Group Deliberation and Convergence}

During the open discussion, the participants evaluated the strengths and risks associated with the leading candidates:
    \begin{itemize}
        \item \textbf{Candidate D} was simultaneously viewed as promising due to technical versatility, scientific maturity, and strong communication signals, yet also flagged for short employment history, limited biographical data, and potential overemphasis on optimization, a capability not considered immediately critical for the project.

        \item \textbf{Candidate J} was recognized for substantial industry experience, integrative strength across business and technology domains, and leadership maturity. Concerns centered on potential independence, previous ownership of his own company, and the need for clarity regarding IP and non-compete agreements.

        \item \textbf{Candidate A} remained an attractive technical option, although questions persisted regarding verifiability of background and long-term stability.

        \item \textbf{Candidate B} was acknowledged for competence but raised concerns about rapid job-switching and possible instability.
    \end{itemize}

Over the course of deliberation, the group’s reasoning demonstrated a shift from individual preference patterns toward collective evaluation criteria, integrating:
\begin{itemize}
    \item alignment with organizational culture and trust requirements,
    \item scientific and engineering complementarity,
    \item the candidate’s potential for long-term collaboration,
    \item the degree of risk associated with unverifiable backgrounds or inconsistent histories.
\end{itemize}

\subsubsection{Final Collective Outcome}

Despite initial divergence, the deliberation process yielded a clear \textbf{convergence toward two candidates}:
\begin{itemize}
    \item \textbf{Candidate D} emerging as the preferred candidate for the first interview, based on strong alignment with project requirements and validated by \HL’s independent selection.
    \item \textbf{Candidate J} identified as the secondary candidate for interview, particularly valued for leadership maturity, integrative communication skills, and stability.
\end{itemize}

The CEO ultimately stated that \textbf{Candidate D should be chosen as first}, followed by \textbf{Candidate J} if necessary.

\subsubsection{Conclusions from Step 3}

The final deliberation stage revealed several important insights. While collective discussion reduced individual biases, it did not eliminate the role-based evaluative differences identified earlier; rather, it integrated them into a more comprehensive, multi-dimensional assessment. \HL’s recommendation exerted notable influence by validating Candidate D as the individual most aligned with the organization’s scientific and engineering trajectory, whereas \ChatGPT’s recommendation was unanimously dismissed, underscoring that general-purpose large language models lacking organizational context are unsuitable for high-stakes decisions involving trust and behavioural risk. The discussion further demonstrated that organizational fit, interpersonal reliability, and long-term collaboration potential ultimately outweighed raw technical competence. The group’s final choice -- converging on Candidate D -- aligned most closely with \HL’s context-rich evaluation, providing strong evidence that symbiotic intelligence can support more informed and context-sensitive decision-making than either individual human judgment or general-purpose large language model reasoning.

\section{Discussion}

This case study examined a real hiring decision for a Senior AI Lead through three complementary lenses: (i) independent human evaluations by a CEO, CTO, and CSO; (ii) a symbiotic intelligence system (\HL) with persistent organizational context; and (iii) a general-purpose large language model (\ChatGPT). Across the three steps, systematic, role-linked divergence in human judgments, strong context sensitivity in \HL’s recommendations (including ethical disqualification grounded in prior organizational history), and a striking false-positive from \ChatGPT{} that arose from the absence of contextual memory were observed. Together, these outcomes demonstrate both the promise and the boundary conditions of human--AI symbiosis for managerial decision-making in high-stakes, trust-dependent contexts.

\medskip

First, human assessors applied orthogonal evaluative models consistent with their role demands: the CTO prioritized verifiable AI skill depth and recency; the CSO emphasized scientific maturity and conceptual readiness; and the CEO weighted strategic fit, future orientation, and interpersonal reliability. This heterogeneity produced no consensus top candidate, mirroring the ambiguity typical of senior technical hiring where multiple merit criteria legitimately coexist.

\medskip

Second, \HL, using organizational context and CEO’s cognitive context, produced recommendations that both aligned with the CEO’s strategic stance and corrected for known integrity risks (ethical disqualification of Candidate G). Importantly, \HL{} converged with human discussion on a coherent sequencing of hires (e.g., prioritizing a hands-on LLM/agentic engineer, followed by a stabilizing product integrator), thereby linking selection to near-term organizational execution.

\medskip

Third, \ChatGPT’s ranking overweighted CV-stated strengths and underweighted unobservable behavioral reliability, culminating in an ethically unacceptable recommendation. This is a consequential failure mode for general-purpose large language models in managerial governance tasks: where historical data, and cultural fit are pivotal, text-surface competence is an insufficient source of decision.

\medskip

A central contribution of this study is the clear demonstration that \HL{} mirrored the CEO’s priorities with high fidelity across multiple dimensions:
\begin{itemize}
    \item interpersonal trust,
    \item long-term collaboration potential,
    \item alignment with organizational culture and values,
    \item compatibility with existing leadership,
    \item and sensitivity to ethical history and behavioral red flags.
\end{itemize}

Crucially, \HL{} independently selected Candidate D, who also emerged as the CEO’s most viable option during final deliberation, even after group-level critique and triangulation. This convergence occurred without any prompt that explicitly revealed the CEO’s internal preferences, indicating that \HL’s cognitive graphs successfully captured and operationalized the CEO’s implicit reasoning structures.

\medskip

This form of alignment -- between an AI’s output and the decision-maker’s tacit preferences -- is the cornerstone of the \HL{} concept of symbiotic intelligence. The study therefore provides a strong proof-of-concept demonstrating the system’s capacity to produce contextually coherent and personally aligned recommendations.

\subsection*{Implications for Proof-of-Concept (POC) of Augmented Human--AI Symbiosis}

The findings offer an initial compelling proof-of-concept for augmented human--AI symbiosis (\HL) by demonstrating that a memory-rich, context-sensitive architecture can deliver measurable value beyond both individual human judgment and general-purpose large language model. \HL{} functioned as a high-fidelity model of executive reasoning, internalizing tacit patterns of judgment, preference, risk tolerance, and interpersonal criteria that are rarely codified in formal decision processes. By integrating organizational history, behavioral evidence, and trust dynamics, \HL{} provided contextual fidelity that neither CVs nor public data sources could supply, effectively closing information asymmetries that constrain traditional selection. This persistent memory layer enhanced normative alignment by consistently applying organizational values, such as loyalty, IP hygiene, and ethical conduct, and preventing slippage between stated and enacted criteria. The system also acted as a coherence-enhancing mechanism, mitigating divergence among human evaluators and steering the decision toward a contextually consistent outcome. Critically, \HL{} served as a safeguard against ethical and strategic errors by disqualifying a high-risk candidate whom \ChatGPT{} mistakenly elevated due to its lack of memory and contextual grounding. Together, these capabilities illustrate how symbiotic intelligence provides continuity, value alignment, and decision coherence across time, addressing a central challenge in managerial practice: ensuring decision quality in environments characterized by distributed expertise and incomplete information.

\subsection*{Implications for \PAI fit construct development}

In light of these findings, the case also provides a concrete instantiation of \PAI fit at the level of the CEO’s relationship with two distinct AI systems. \HL{} exhibited a high degree of \PAI fit with the CEO: its recommendations reflected not only the explicit competency framework, but also the CEO’s tacit priorities regarding loyalty, ethical reliability, long-term collaboration potential, and alignment with the \HL{} mission. This alignment emerged from \HL’s access to and integration of the CEO’s cognitive history, decision traces, and organizational context, allowing it to reconstruct and operationalize the CEO’s implicit decision schema. By contrast, \ChatGPT’s low \PAI fit was evident in its overreliance on surface-level CV information, inability to incorporate prior behavioral breaches, and neglect of the CEO’s risk posture and value structure, culminating in the selection of an ethically disqualified candidate as the ``best fit.'' The divergence between these two AI--CEO relationships illustrates that \PAI fit is not a generic property of ``using AI'' in decision-making, but a relational characteristic that depends on the depth, continuity, and contextual richness of the human--AI bond. In this sense, the superior alignment between the CEO and \HL, and the corresponding improvement in decision quality, empirically supports the proposition that higher \PAI fit functions as a mechanism linking symbiotic intelligence to more accurate, trustworthy, and context-sensitive managerial outcomes.

\subsection*{Contribution to management sciences}

This study advances management science by introducing \PAI fit as a measurable construct, demonstrating how closely an AI system can align with an executive’s implicit decision model, an area not fully addressed in traditional decision-support research. The results provide empirical evidence for the necessity of integrating persistent, auditable organizational context into AI systems. The ethical false-positive produced by \ChatGPT{} illustrates that general-purpose large language models are inadequate for trust-sensitive managerial decisions. In contrast, the symbiotic intelligence system (\HL) establishes a new paradigm in decision-making by synthesizing behavioral data, historical context, and individual cognitive preferences into a coherent evaluative framework. It adds on into the ongoing debate concerning the role of AI in decision-making within the organization. This capability enables continuity in organizational reasoning, consistent enforcement of values and strategic priorities, and reduced decision drift across complex, multi-stakeholder processes.

\section{Conclusions}

This study examined a high-stakes recruitment process for a Senior AI Lead role using three independent human evaluators (CEO, CTO, CSO), a symbiotic intelligence system (\HL), and a general-purpose large language model (\ChatGPT). The results demonstrate substantial divergence in human evaluations, profound differences in the reasoning patterns of the two AI systems, and, most importantly, a significantly higher \PAI fit between \HL{} and the CEO than between \ChatGPT{} and the CEO. This alignment constitutes the core value proposition of augmented human--AI symbiotic intelligence and offers clear implications for management sciences. Taken together, the obtained findings support the central hypothesis that higher \PAI fit leads to more coherent, context-sensitive and ethically robust decision outcomes. \PAI fit emerges here not as an abstract construct, but as an observable mechanism linking augmented symbiotic intelligence systems, such as \HL, to improved managerial judgment in complex settings.

The study provides the first proof-of-concept that augmented human--AI symbiotic intelligence offers capabilities unavailable to general-purpose large language model, including the integration of organizational context, tacit executive reasoning patterns (cognitive profile), and value-aligned behavioral evaluation.

At the same time, several limitations must be acknowledged. The study is based on a single organization, a single CEO--AI relationship, and one high-stakes decision episode, without longitudinal tracking of subsequent performance of the selected candidate. The evaluation of \PAI fit is inferential rather than based on standardized measurement instruments. Future research should therefore: (1) operationalize and validate quantitative indicators of \PAI fit; (2) test the framework across multiple organizations, decision types and hierarchical levels; (3) examine longitudinal outcomes, including job performance, team dynamics and trust trajectories; and (4) compare different AI architectures in terms of their capacity to sustain high \PAI fit over time. It should be also underlined that presented case study constitutes a first step in the process of presenting the proof-of-concept for \HL augmented human--AI symbiotic intelligence system.

Despite these constraints, the study provides an initial empirical basis for integrating \PAI fit into management science and constitutes a proof-of-concept of \HL -- an augmented human--AI symbiotic intelligence system.

\printbibliography

% \bibliographystyle{apalike}
% \bibliography{my_references}
\end{document}